# A Model of Morphogenesis

F. W. Cummings [1]

## Abstract

A new model is presented of a patterning mechanism in animals. It is argued that the 'Wnt' signaling pathway often acts in binary conjunction with a variety of other pathways to produce complex patterning. This patterning follows while involving little of the detailed biochemistry of either of the two pathways involved. The patterns arise spontaneously from a state of zero pattern activation upon achieving a specified (epithelial) sphere radius. Gastrulation follows from the simplest solution, and assuming a simplest form for the coupling of the Gauss and Mean curvatures to the morphogens. Numerical solutions are presented, and predictions of the model are discussed.

[1]Present address: 2365 Virginia St., Berkeley, Ca. 94709

email: "fredcmgs@uclink.berkeley.edu

## 1. Introduction; Patterning and Morphogens

Morphogens are secreted signalling molecules that provide patterning for a field of surrounding cells. One well-known and widely discussed hypothesis is of a gradient of concentration emanating from a localized source, which determines the fate of cells according to the amplitude or concentration of the morphogen (Wolpert 1989; Wolpert 1998; Gurdon & Bourillot 2001). In this picture, cells interpret a concentration gradient. An alternate hypothesis for the establishment of patterns in developing organisms was introduced by Turing (Turing 1952). The 'reaction-diffusion' mechanism, while sharing some similarities with the first mentioned mechanism, has significant differences. A minimum of two morphogens is of course required for a reacting system, and the patterns arise autonomously without reference to a source. Nonlinearity is an essential ingredient in the latter model and is the source of the instability (Turing 1952; Koch & Meinhardt 1994). Also, in Turing type models an essential ingredient is the existence of two diffusion constants of disparate values. A more recent interesting approach to patterning is based on interacting gene networks (Von Dassow et. al. 2000; Salazar-Ciudad, Newman & Sole 2001). These models contain ~ fifty parameters and give a segmentation pattern. An interesting model of morphogenesis has been given recently by Beloussov (2003).

The present work develops a simple alternate conceptual model for the establishment of morphogenetic patterns, one based on a pair of interacting signalling pathways, and referred to here as the "ISP" (interacting signaling pathways) model. The emphasis here is on the straightforward way in which patterns extending over many cell diameters may emerge from coupled signalling pathways. Rather than dealing with the complexities of the signalling pathways, only those necessary components involved in the patterning need be included.

The present model differs conceptually in two important respects from the well-known Turing type model. First of all, pattern formation arises spontaneously in the ISP without invoking the essential role of nonlinearities, although nonlinearities certainly and importantly enter; nonlinearities in the reaction terms of the 'reaction-diffusion' conceptualization in Turing models are the source of spontaneous pattern formation. Secondly, the two ligands in the present model diffuse only one or a few cell diameters, and their diffusion constants are not required to be different. After fifty years it cannot be said that there is good evidence in a developing biological organism for a 'reaction-diffusion' mechanism where two 'morphogens' with different diffusion constants travels over many cell diameters and form developmentally relevant patterns. The present model is then conceptually very different from such 'reaction-diffusion' mechanisms, and, it is believed, conceptually simpler and more realistic, being based on signaling pathways. The patterning achieved is in many, perhaps most cases, very similar in the two models, even though the conceptual basis is here very different.

There have been a number of previous works that obtain patterns by modeling signaling pathways, and especially notable among these are those of the Delta-Notch system. These latter give patterns on a much smaller spatial scale than those of the present model, and may be viewed as possibly acting in conjunction with the present pattern mechanism, coming into play subsequently to the ISP (Bosenberg and Massague 1993; Collier et. al. 1996; Lewis 1998; Owen, Sherratt & Wearing 2000; Sekimura et. al. 1999).

In section 3, plausible signaling pathways are discussed as stand-ins for the symbols of the model. The principal one is the 'Wnt' pathway, in conjunction with associated G proteins. Only the essential ingredients necessary to establish the pattern are considered; the model is formulated with as little extraneous detail so as to emphasize the crucial elements. Further, the intent is to give a model of morphogenesis, as contrasted with simple patterning, so that in particular the signalling pathways are envisioned as those most closely coupled to and influencing such key early developmental events as cell architecture (and thus epithelial sheet folding), adhesion, differentiation and cell division (Miller et. al. 1999; Gerhart & Kirschner 1997; Raff 1997).

The elements of the model are two (free) extracellular ligands, with densities (# per unit area) $L_1$ and $L_2$, and the densities of their activated pathways (or simply their activated receptors) $R_1$ and $R_2$. The latter two consist of a combination of a receptor with its associated ligand. The ligands are supposed to diffuse in the extracellular space, and the diffusion length need be only as large as a few cell sizes, but not necessarily so small.

There are two simple elements of the model. Attention is focused on a small cluster of cells, when use of such terms as 'ligand density' and 'receptor density' has meaning. These cells are to be thought of as comprising a closed epithelial surface, so that the densities of the model are 'numbers per unit area'. Variation of the morphogens (the R's or L's) along the apical-basal direction is not considered, or rather thought of as being an averaged value in this dimension. First of all, each such 'cell' produces ligand of like kind proportional to the level of receptor activation; $R_1$ stimulates production of $L_1$, otherwise the process would be limited to a purely local one if "like-ligand" production were not induced, and the particular cell in question would act as a 'sink'. The second key element in the model is that each activated pathway acts to inactivate the other; as $R_1$ increases, the level of ligand production $L_2$ is decreased, and similarly for $R_2$.

The equations representing such a process are

$$\frac{\partial}{\partial t} L_1 = D_1 \nabla^2 L_1 + \alpha R_1 - \beta R_2 + NL., \qquad 1)$$

$$\frac{\partial}{\partial t} L_2 = D_2 \nabla^2 L_2 + \beta R_2 - \alpha R_1 + NL., \qquad 2)$$

$$\frac{\partial}{\partial t} R_1 = C_1 \overline{R}_1 L_1 - \mu R_1, \qquad 3)$$

$$\frac{\partial}{\partial t} R_2 = C_2 \overline{R}_2 L_2 - \nu R_2. \qquad 4)$$

The first two terms in eqs.1), 2) represent in the usual way diffusion of the ligands in the extracellular space. All parameters in the model (e.g., $\alpha$, $\beta$, $D_1$, $\mu$, $\nu$) have positive values, as do also, of course, the densities $L_1$, $L_2$, $R_1$ and $R_2$. The terms $\alpha R_1$ in eq.1) and $\beta R_2$ in eq.2) represent the production of 'like' ligand by the corresponding activated receptor. These same terms are used to represent the fact that activation of receptors of density $R_2$ deactivate or turn off production of free ligands of density $L_1$, and vice versa. The system acts as a 'toggle'; a region of high activation of one implies low activation of the second. The term NL on the r/h/s of eqs.1) and 2) indicate that there are expected to be nonlinear additions to the toggle; saturation will set in for large enough values of either active receptor density. The transmembrane receptors, which reside in the lateral cell membrane, are relatively immobile. The respective activated densities decay at rates $\mu$ and $\nu$, and this 'decay' returns the receptors to their inactive state. Two first terms on the right side of eqs.3), 4) say that there is a positive rate of change of $R_1$ or $R_2$ proportional to both the density of empty receptor sites ($\overline{R}_1, \overline{R}_2$) and also to the density of free ligands at the particular local cell site. The density of empty sites may be obtained from the expression

$$R_1 + \overline{R}_1 = R_o + \eta R_1, \qquad (R_0 = const.),$$

where the last term on the r/h/s expresses the possibility that the total number of receptors of each type (e.g., '1') increases with activation of that same type receptor, and new (empty) receptors are thus added. Then the empty receptor site density may be written

$$\bar{R}_1 = R_o(1-\varepsilon_1 R_1), \qquad (\,0 < \varepsilon_1 \leq 1, \varepsilon_1 \equiv (1-\eta)/R_o\,), \quad 5)$$

and similarly for type '2'. The values $\varepsilon = 1$ (and $\eta = 0$), implies that there is no receptor augmentation $\sim R_1$, while $\varepsilon \sim 0$ implies either that there is a new empty receptor created for (almost) every one occupied, or that there are very many more empty sites than occupied ones. When eq.5) and the analogous equation for type '2' is used in eqs.3), 4) to eliminate the unoccupied site densities, the model then comprises four coupled equations for four unknowns. The coupling from epithelial shape to morphogen, and back, has not as yet been specified (cf. section 4).

The process of producing ligand upon activation of the cell surface receptor could possibly involve numerous steps, involving (e.g.) gene transcription, the endoplasmic reticulum (ER), the Golgi complex, and finally secretion from the cell. This time is expected to be considerable compared to the time for a free ligand in a given spatial region to become attached to its receptor and to activate the pathway. However, it is supposed here instead that R acts to release already stored ligand, (stored at, e.g., a constant rate by an unspecified cellular mechanism). The cell maintains a relatively constant store of ligand awaiting a release signal $\sim R$ (analogous (in this respect only!) to the situation of neurotransmitters in neurons). The two times (emission time of new ligand $\sim R$ and empty receptor uptake of ligand L) can thus be comparable. This is the situation envisioned here, and will have to serve as a prediction of the model at this point: the activated receptor $R_{1,2}$ releases ligand already stored in vesicles, so that this time is appreciably shorter than ligand production and storage via gene activation, ER and Golgi. Importantly for the model, Wnt (cf. sec. 3) has two known modes of action, one that bypasses the nucleus, and a second 'canonical' pathway leading to gene activation via stabilization of beta-catenin. The former 'non-canonical' path bypassing the nucleus acts (at least in part) to release stored Wnt ligand relatively rapidly.

Wnt has recently been purified (Willert et. al., 2003), so that there is the possibility that the prediction that receptor activation by Wnt ligand leads to relatively rapid stimulated secretion of further Wnt ligand can be checked experimentally, perhaps by the use of GFP (green florescent protein) as a marker for the inducing ligand. This time difference between stimulation and release should be significantly shorter than the time expected if the process occurred by way of transcription.

The particular solution picked out to eqns.1-4) will be determined by boundary conditions. Most generally, the boundary conditions are expected to be zero slope conditions. Imagine as the simplest example a hollow sphere of some surface thickness, and a one dimensional pattern on the middle surface of this sphere of the morphogen $R_1(u)$, where u is the coordinate which runs from the north to south pole. Then requiring continuous slope at the north and south poles requires zero derivatives of R at these two points. When this is the case for the morphogens, then the solution to the Laplace equation of the model (eq.10 below) will be a constant.

2. Stability of the Morphogens near Zero Activation

An important requirement of any model is that it gives spontaneous pattern formation when the shape reaches a certain critical minimum size. In particular, as will be seen on hand of the present model, such a critical size provides a point of onset of patterning leading to gastrulation when the blastula reaches a specified radius (of the middle surface). A Hydra segment, for example, below a certain size will not regenerate. This is a criterion given in terms of the parameters of the model, and one not previously given.

In order to find the point at which a pattern spontaneously begins to emerge in the present model, one considers the small amplitude (linear) version of the model equations, and sets all four variables of the ISP in section 1 proportional to Exp(-s·t+k·x) in the usual way (Murray 1990), where 't' is time and 'x' is space. A fourth order polynomial equation is obtained from eqs. 1→4) for the variable 's', and the condition s < 0 is sought, when a pattern will emerge spontaneously. It turns out straightforwardly that s < 0 when the condition $k^2 < k_c^2 \equiv k_1^2 + k_2^2$ is satisfied, where $k_1$ and $k_2$ (~ length$^{-1}$) are defined by the parameters of the model by

$$k_1^2 = \alpha C_1 R_o/(D_1\mu), \quad \text{and} \quad k_2^2 = \beta C_2 R_o/(D_2\nu). \quad 6)$$

This is easily seen by setting s = 0 in the fourth order polynomial F(s) below and requiring that the resultant expression F(0) be less than zero, which is sufficient to ensure that at least one root will be negative. The parameters of the model are such as to give only one real negative root $s_o$ in parameter ranges of most interest. The dimensionless variable F(S) is

$$F(S) = (S-x^2d)\cdot(S-x^2/d)\cdot(S-m)\cdot(S-1/m) - p\cdot(S-x^2d)\cdot(S-m) - q\cdot(S-x^2/d)\cdot(S-1/m). \quad 7)$$

In eq.7) the definitions of the relevant dimensionless quantities S, x, d, m, p and q are

a) $S = s/\sqrt{\mu\nu}$;  b) $x^2 = k^2\cdot\sqrt{D_1D_2}/(\mu\nu) \equiv k^2/k_o^2$;  c) $d = \sqrt{D_1/D_2}$;

d) $m = \sqrt{\mu/\nu}$;  e) $p = \beta C_2 R_o/(\mu\nu)$;  f) $q = \alpha C_1 R_o/(\mu\nu)$.

8)

It is also helpful to rewrite eq.7) in a form so that F(S) = 0 implies that

$$G(S) = p\cdot q, \quad \text{or} \quad ((S-x^2d)\cdot(S-m)-p)\cdot((S-x^2/d)\cdot(S-1/m)-q) = p\cdot q. \quad 7')$$

Now since each of the two factors in eq.7') is a quadratic, we can obtain the four roots of the left hand side quartic G(S) at once. They are all real. Then, upon graphing a quartic G with four real roots, we look for its intersection with the constant curve p·q, and in particular ask when this intersection has a negative root $S_o$. Clearly then, this occur at the maximum value of k (~ minimum animal size) such that there is an intersection of the quartic G of eq.7') with the S = 0 axis at a point just below intersection of G with the constant curve p·q. As x (or

k) decreases, this root becomes increasingly negative, as shown in Figure 1.
There are two imaginary roots S of the function $F = G - p \cdot q$ when the maximum of G is below the constant curve $p \cdot q$. There are always at least two real roots, and one of these is negative in the appropriate range where $k^2 < k_C^2$ (or $x^2 < x_C^2 = q/(md)+p \cdot (md)$). The minimum size then corresponds to a 'length' $2\pi/k_C$, and no pattern is possible for a smaller length. Figure 1 shows a typical plot of the four roots $s_i$ of the equation $F(s) = 0$, shown as a function of the (dimensionless) 'wave vector' $k/k_o$. It is of interest to note that the other two roots can be complex in general, as the imaginary part provides for oscillating (and exponentially dying) solutions.

For sufficiently small amplitudes, the linear steady state equations are

$$\nabla^2 (R_1 - fR_2) + (k_1^2 + k_2^2)(R_1 - fR_2) = 0, \qquad 9)$$

and

$$\nabla^2 (R_1 / k_1^2 + fR_2 / k_2^2) = 0. \qquad (f \equiv \beta/\alpha).$$

10)

It is seen that the variable $(R_1 - f \cdot R_2)$ obeys the well-known Helmholtz equation (c.f., e.g., Murray 1990), while the variable $(R_1/k_1^2 + f \cdot R_2/k_2^2)$ obeys the equally famous Laplace equation. Murray (1990) has shown that many suggestive animal coat patterns arise from the Helmholtz (or 'time-independent wave') equation. The variety of patterns obtainable from the above equations is extremely rich, and any given pattern depends critically on the geometry as well as the boundary conditions. It is to be noticed that upon inclusion of the indicated nonlinear terms, the discrete eigenvalues nature of solutions no longer obtains, and while the form of the patterns are generally maintained, the amplitude of the R's will increase with increasing size or area. A limiting size is reached when the R's can no longer remain positive with the given boundary conditions, whereupon the new solution must become the next allowed low amplitude and more complex one.

It is also of some interest that the linear (small amplitude) version of the model eq.9) already gives the largest majority of plant patterns, namely, the spiral phyllotaxis specified by the Fibonacci series, as well as the 'whorled' phyllotaxis, and also the bijugate, trijugate, … (Cummings & Strickland 1998). It is interesting to note that there is no a priori structural reason for the observed very low relative rate of occurrence of four leaves on a stem, as compared to two, three or five in spiral phyllotaxis. Natural selection of course acts in choosing the adaptive patterns, but clearly does not address the relative uncommonness of a spiral 'four' pattern, remembering that a 'three' or 'five' Fibonacci spiral is common. This perhaps indicates that there is an analogue in plants to the signaling pathway patterning mechanism of the present model, although the plant biochemicals involved are expected to be quite different than the animal ones (Meyerowitz 2002).

Figure 2 shows a typical numerical solution of the ISP for the two-receptor densities, as well as the derivatives, in the steady state (when $\partial/\partial t = 0$), on a cylindrical-like axially symmetric shape in one (dimensionless) space coordinate 'u'. Complex patterns are expected from this model as can be seen when it is realized that the linear version of the model in the steady state gives the well known associated Legendre patterns on a sphere (see Murray 1990, p.398), and generally gives periodic behavior in one dimension. When nonlinearities are included, eigensolutions no longer exist, and the patterns change as the sphere radius increases, and the amplitudes of the solutions increase as the total area 'A' increases. The ISP contains an original handful of parameters, and these group into only four significant ones. (excluding parameters (~ 2–3) which enter into the connection between the epithelial surface curvatures and morphogens in section 4).

3. Biochemical Interpretation for the Model

Here are reviewed recent findings to suggest elements of two different signaling pathways as prime candidates that may be best represented by the symbols of the model. Much evidence suggests that one symbol pair of the model (say, $L_1$, $R_1$) represents the 'Wnt' (wingless, 'Wg', in the fly) pathway. It is suggested that the Wnt pathway is coupled to any one of a number of possible other candidates. One example is the interaction between Wnt and Vg1 signaling pathways initiating primitive streak formation in early chick embryos (Skromme, I. & Stern, C., 2001), while a second is that between Wnt and Hedgehog, affecting segmentation in Drosophila (Gerhart and Kerschner, 1997). Yet another example is the combination of a dorsal signal provided by the BMP4 homologue Decapentaplegic (Dpp) with a ventral signal provided by the Wnt homologue Wingless (Wg) establishing the PD (proximal-distal) axis, in addition to organizing the dorsal-ventral appendage pattern (Galindo et.al, 2002). Evidence from other arthropods and vertebrates suggest that this PD patterning mechanism is probably conserved and ancestral. Epithelial bud development is due to the combined action of Wnt and BMP (bone morphogenetic protein) signaling pathways. In this case, the β-catenin activated Lef1 transcription complex is combined with the 'noggin' inhibition of BMP signaling, leading to reduced production of the adhesion protein E-cadherin, the latter mediating cell-cell contacts (Jamora et.al, 2003; Perez-Moreno, Jamora & Fuchs, 2003). Another most interesting example of Wnt signaling (and Wnt6 in particular) is that given by Bonner-Fraser and colleagues. Neural crest cells are a pleuripotent, migratory population of cells that differentiate into an enormous array of cell types, tissues and organs in vertebrates. Wnt6 is both necessary and sufficient for instructing neural crest formation in the chick embryo (Garcia-Castro, Marcelle & Bonner-Fraser, 2002). The present ISP model suggests that it will turn out that a synergy between BMP and Wnt is required for neural crest formation. A further example is the coupling of Wnt to the Notch pathway, as in animal segmentation.

A molecular oscillator underlies vertebrate segmentation, and involves the interaction of Wnt with the Notch pathway (Purnell, 2003; Pourquie, 2003). Apparently the Notch pathway lies at the heart of the vertebrate oscillator, although there has been one cycling Wnt gene uncovered also, namely the Wnt inhibitor gene axin2. Notch is not involved in Drosophila segmentation, although involved in other arthropods, suggesting that vertebrates and arthropods may have shared a common ancestral segmentation program with Wnt/Notch signaling at the core, and parts of this program was lost in particular descendant lineages such as Drosophila (Stollewerk, Schoppmeir & Damen, 2003).

It also seems possible that, besides the possibilities mention just above, there may be more than one version of the Wnt protein, since Wnt is known to associate with lipids and become hydrophobic. Then it may be possibly think of Wnt1 and Wnt2 in this case as the two ligands of the model (Willert et. al., 2003).

Signaling pathways are accompanied with their modulating 'G' proteins (GTP and GDP), the G proteins playing the role of the 'switches' of the model. To use a mechanical analogy, the Wnt pathway seems to act as a sort of permissive gate, perhaps a sort of 'main drive shaft' off of which other 'shafts' are driven or interact. The 'Wnt' pathway may be looked on as a sort of primal pathway, participating along with numerous other pathways (e.g., Hedgehog and Notch (Gerhart & Kirschner 1997; Cooper 1997)) in early development of an organism. Cell signalling via the Wnt 'Frizzled' receptor has evolved to considerable complexity within the metazoans. The Frizzled–dependent signalling cascade comprises several branches, whose differential activation depends on specific Wnt ligands, Frizzled receptor isoforms as well as the cellular context (Moon & Shah 2002; Niehrs 2001; Tetsu & McCormick 1999; Taipale & Beachy 1998; Wodarz & Nusse 1998). For example, in Xenopus laevis embryos, the canonical β-catenin pathway contributes in a crucial way to the establishment of the dorsal-ventral axis, involving interaction between the Wnt/β-catenin 'canonical' branch and a Wnt/$Ca^{2+}$ branch. Mutations in the protein APC, a key regulator in the Wnt pathway, leads to accumulation of β-catenin, which in turn activates genes which respond to transcription factors of the TCF/LEF family, with which β-catenin interacts (Tetsu & McCormick 1999). Members of the large Wnt family of proteins control many developmental processes, including a pathway involving cell adhesion (Niehrs 2001; Wodarz & Nusse, 1998). This pathway is a crucial factor in construction of the colon (Peifer, 2002). Of most interest from the present point of view of the model is the non-canonical action, a pathway which bypasses the nucleus altogether.

It is proposed that the 'switch' activity of the present model is likely provided by the small guanosine triphosphatases (GTPases, and especially Rho), and is modeled by the R terms in eqs.1) and 2). GTPases are molecular switches that use a simple biochemical strategy to control complex cellular processes. They cycle between two conformational states; one bound to GTP (the 'active state'), the other bound to GDP ('inactive state'), and they hydrolyze GTP to GDP. In the 'on', or GTP state, GTPases recognize target proteins and generate a response until GTP hydrolysis returns the switch to the 'off' state. This idea has been elaborated throughout evolution, with a mammalian cell containing several hundred GTPase switches. The Ras superfamily of GTPases are master regulators of numerous aspects of cell behavior. These small monomeric GTPases fall into five major groups: Ras, Rho, Rab, Arf and Ran. Rho GTPases are an important example (Etienne-Manneville & Hall, 2002). They participate most importantly in the regulation of cell polarity. Typical epithelial cells, such as those of interest in the present model, and such as (e.g.) those lining the colon and also forming the Drosophila wing, form monolayers of packed cuboidal cells with specialized cell-cell contacts and a distinctive asymmetric distribution of proteins at the basolateral and apical membranes. It is these asymmetrically distributed proteins which determine the cell shape, and thus the epithelial surface shape, and are then most interesting from the point of view of the model here. The GTPases of course have their regulators. A most important example, Moesin, acts antagonistically to the Rho pathway to maintain epithelial integrity. In Drosophila, Moesin functions to promote cortical actin assembly and apical-basal polarity. Cells lacking Moesin lose their epithelial character and adopt invasive migratory behavior. When Moesin is mutated in the fly imaginal epithelium, a single layer of tall columnar cells with an obvious apical basal polarity, their cells lose intercellular junctions and epithelial polarity, and are extruded basally from the epithelium. Apparently Moesin facilitates epithelial morphology not by providing an essential structural function but rather by antagonizing activity of the small GTPase Rho, thus regulating cell-signaling events that affect actin organization and polarity (Speck et. al., 2003).

Studies of Xenopus embryos and cultured cells show that Wnt/Frizzled signaling activates the cytoskeleton regulator Rho through activation of Dishevelled. Dishevelled is a multifunctional protein that regulates cell polarity through the non-canonical pathway, but also regulates cell fate through the canonical Wnt/beta-catenin pathway. These results link Wnt/Frizzled signaling during epithelial sheet movement to the Rho family of GTPases. (R. Keller, 2002). The non canonical Wnt/Frizzed pathway leading to production of further Wnt ligand (represented by the '+' terms in eqs.1),2)) seems by best present guess to go by way of Wnt/Frizzled $\rightarrow$ G protein $\rightarrow$ Ca++ $\rightarrow$ PKC $\rightarrow$ Cdc42 $\rightarrow$ ? (Keller, 2002, fig.4; Harwood & Braga, 2003). The conjecture is that this is the pathway that turns off the release of '2' ligand at the same time that it stimulates release of 'same' ligand $L_1$, and symbolized by the r/h/s of eqs.1) & 2). A similar '2' pathway is conjectured to operate in an analogous way to turn off release of ligand '1' (Wnt), at the same time as stimulating release of the already stored ligand '2'. Such a pathway has as not been described, but is predicted by the present model.

The density $R_1$ of the model then represents a 'Wnt+Frizzled' (ligand + receptor) combination, while $L_1$ represents unbound ('free') Wnt ligand. Then the '2' subscript represents the second pathway coupled to Wnt, e.g., the Hedgehog (Hh), BMP or Notch pathway (or homologues).

### 4. Gastrulation

A speculation is that discovery by evolution of this simple coupling of two already existing signaling pathways could possibly have been a key event at or near the Cambrian boundary leading to the abrupt transition from single celled organisms to multicellulars, esp. those animals with definite patterning and defined, polarized epithelial cell sheets. Of course the setting must have been accommodating also, e.g., complex eukaryotes had evolved over one to two billion years, the oxygen level was adequate, and an initial fierce competition was lacking. Arguments as to the adaptive value of such a coupling between signaling pathways, one giving patterned epithelial closed surfaces, will be more apparent if the case is made for gastrulation occurring in a straightforward manner from an initial hollow ball of cells.

All animals have two or three cell layers, ectoderm, endoderm and mesoderm. Gastrulation is the most fundamental and earliest epithelial folding process in any animal's life, and is the process of producing an inner layer (endoderm) and an outer layer (ectoderm). These two layers are the primal 'gut' or digestive zone, and the outer protective layer. Most animals have a third type called mesoderm comprising the 'middle' organs between the last mentioned, giving rise to such organs as heart, lungs, kidneys, etc.

Evolution of a 'proto' digestive zone offers an apparent advantage in the struggle for survival, since such a relatively large animal can devour many single celled animals and plants at a gulp. At the same time, it is realized that formation of a patterned clump of cells is clearly not adaptive 'per se'; such must at the same time be adept at catching and devouring prey. The following points out that gastrulation follows easily from the patterning scheme laid out above, with a couple of extra key assumptions. We need to ask next how the morphogens affect the surface curvatures, and vice versa.

For purposes of visualization, it is helpful to imagine for the moment a 'cell' (or rather, a small collection of actual cells) as a deformable 'can', a right cylindrical figure with height 'h'. It can be visualized that one of the morphogens acts to constrict the apical area at the same time as increasing the basal area, while the action of the second morphogen is the opposite. In mathematical terms, we assume that we may express the Mean curvature of the thick surface at any point by the expression

$$H = \sqrt{\frac{4\pi}{Area}} \cdot \left(1 + \lambda_1 \left(R_1 - R_2\right)\right). \qquad 11)$$

Here $\lambda_1$ is a constant and 'Area' indicates the total (closed, middle) surface area, and $R_1$, $R_2$ are here taken as (dimensionless) morphogens.

This form for the Mean curvature arises from the consideration that the surface will remain a hollow sphere (the shape of minimum adhesive energy) as long as the size is too small (radius = $R < R_o$) to admit a solution ($R_1, R_2 \not\equiv 0$) to the model equations, in particular to the Helmholtz equation, in which case $H = 1/R$. Here we will assume steady state, $\partial/\partial t = 0$, while the much slower time dependence will be carried by the total (increasing) area, $\equiv A$. While a sphere with a radius $R < R_o$, no pattern can exist, and $A = 4\pi R^2$. Then eqn.11) gives the answer that $H = 1/R$ for any $R<R_o$, as required. A small amplitude pattern only begins to emerge when $k^2R^2 \geq 2$, as is well known, and this gives the critical size $R_o$ of the blastula in terms of the parameters of the model, $k^2 = k_1^2 + k_2^2$. In the present discussion, it is assumed that the epithelial sheet thickness 'h' is constant, although this is not a necessary condition in general (Cummings, 2001).

The surface is uniquely described only when two curvatures are given (do Carmo, 1976), and the second, besides H of eq.11), is the Gauss curvature 'K'. From their definitions as the average ('H') and product ('K') of the two principal curvatures in orthogonal directions, it is clear that $K \leq H^2$, so that we may write $K = H^2 - D^2$, with $D = 0$ for a sphere, or in any local region where the surface is 'sphere-like'. In the case of axial symmetry, the case of present interest, we see that $D \to 0$ at the two 'poles' on the axes, where the morphogens must have zero slope. The function 'D' is simply the difference (divided by 2) of the principal curvatures. At the same time, K is an invariant, independent of coordinates, as are the morphogen functions. The three constraints of a) invariance, b) positive definiteness, and c) $D \to 0$ at boundaries, strongly limits the possible form of D, and suggests taking K of the form

$$K = H^2 - \lambda_2 \left( \vec{\nabla}(R_1 - R_2) \right)^2. \qquad 12)$$

Clearly both H and K have been taken of the simplest form, and complications and further parameters can sensibly be added. Here $\lambda_2$ is a constant parameter.

The morphogens affect the geometry as in eqns.11) and 12), but how does the surface geometry affect the morphogens? Circa 1860, Gauss gave the equation (see Cummings, 2001)

$$\nabla^2 \ln(g(u,v)) = -2 \cdot K \qquad 13)$$

connecting the surface metric coefficient 'g' to the Gauss curvature K in a most nonlinear way. A sphere, for example, in one space coordinate 'u' has a most simple metric of $g = R^2 \cdot (1/\cosh(u))^2$, when $K=1/R^2$; this g provides a covering for the sphere excluding only two points, at the north and south poles. Now in the conformal coordinates which we have employed in eq.13), where any surface is always and everywhere divided into small (infinitesimal) squares, the Laplacian takes the form in terms of the coordinates u, v

$$\nabla^2 = \frac{\left(\dfrac{\partial^2}{\partial u^2} + \dfrac{\partial^2}{\partial v^2}\right)}{g}. \qquad 14)$$

The metric in these special (and convenient) coordinates is $\overset{\leftrightarrow}{g} = g(u,v)\overset{\leftrightarrow}{I}$, that is, it is the unit 2×2 matrix times g. Then it is remembered that g alone has no physical interpretation, but only in combination with other quantities, such as (e.g.) the invariant path-length $(ds)^2 = g \cdot (du^2 + dv^2)$, the invariant Laplacian operator $\nabla^2$, or the squared gradient operator $(\nabla f)^2$ in eq.12), and most importantly, the element of area $dA = g \cdot du dv$. It is useful to define a 'shape' function 'G', whose integral is unity, i.e., $g \equiv A \cdot G(u,v)$ where A is the total area. Equation 14) serves to show how the geometry, represented by 'g', affects the morphogens, since the Laplacian operator $\nabla^2$ enters the equations eqs.1)-2) determining the morphogens.

The eqs.1)-5), together with eqs.11)-14) constitute a set of three coupled second order differential equations. These have been numerically integrated in the case of axial symmetry and in steady state ($\partial/\partial t = 0$), when the single coordinate 'u' runs from 'u' = $+\infty$ (~ +10) at one 'pole' and $-\infty$ (~ –10) at the opposite 'pole'. Sample solutions are shown in Figures 4 and 5. (The numerical integrations were carried out with student 'MatLab', and the program is available by email request).

Of particular note is that it is required that the famous (and remarkable) Gauss-Bonnet theorem be obeyed at all times. This theorem says that the integral of the Gauss curvature K over the entire area be a constant = $4\pi n$, where n is an integer or zero. For a 'donut' shape, n = 0, while for any surface topologically equivalent to a sphere, such as the gastrula, n = 1. It is noticed that the region of the surface in the case of gastrulation where the mean curvature H changes sign is necessarily a region of negative Gauss curvature, which may be inferred by consideration of the Gauss curvature at the gastrula 'cusp', where it takes on a 'half-donut' shape; at the cusp, K goes to zero, and the point where K is most negative is where H ~ 0. For the solutions shown, $\lambda_1 = 7$ and $\lambda_2 = 7$, but solutions are available for a range of these two parameters, e.g., $\lambda_1 = 10$ and $\lambda_2 = 5$

## 5. Discussion

The present work presents a morphogenetic pattern algorithm based on the simplest of notions. The key assumption is that activation of one signaling pathway acts to suppress a second, and vice versa. Further details of the signaling pathways are not necessary to establish pattern formation. Arguments are presented that one of these pathways is expected to be the 'Wnt' pathway. This Wnt pathway is coupled to any one of several other possible signaling pathways involved in early development, e.g. Hedgehog (Hh), BMP, Notch (and even possibly to a second 'Wnt' pathway). Such coupling leads easily to complex patterning, involving cell differentiation and epithelial sheet deformations. These pathways are those most crucially affecting cell architecture, the cytoskeleton, adhesion, cell division and cell death in early development. It is a speculation that this discovery of patterning by evolution may have been one of the crucial steps along the path to multicellularity.

The present model is consistent with the commonly seen periodic segmentation patterns, which in both arthropods as well as vertebrates involve both Wnt and Notch. This may indicate a common ancestral origin, while particular later arthropods (e.g., Drosophila) may have evolved a somewhat different mechanism, nevertheless involving Wnt in a key role. Within the complex Drosophila segmentation pattern it is worth noting the apposition of Hedgehog and Wingless (Wnt) expression stripes which appear (of course) after cellularization. Then Wnt is the one invariant patterning pathway, acting in conjunction with an alternating second pathway, with Notch as the most frequent second partner in segmentation. Different partnering of second signaling pathways with Wnt may occur sequentially. Thus it is a prediction that Wnt will always be one of the partners when two signaling pathways act together to produce segmented patterns.

A second prediction of the model, mentioned earlier, is that Wnt ligand activation of the Frizzled receptor will induce further production of Wnt ligand, and in a time short compared to the time for transcription and translation.

Morphogenetic shape transformations are coupled to a plausible pattern formation mechanism, and pattern formation is in turn affected by the geometry of the epithelial surface (Cummings, 2001). Three coupled second order differential equations result. All present numerical solutions are given in axial symmetry, although this is not a necessary restriction of the model.

It is particularly to be noticed that the onset of gastrulation from an original hollow sphere of epithelial cells is shown to be a consequence of the model, so that the fact that there is a spontaneous morphogenetic pattern arising only upon achievement of a finite size of the (middle) sphere radius is a noteworthy result of the present model.

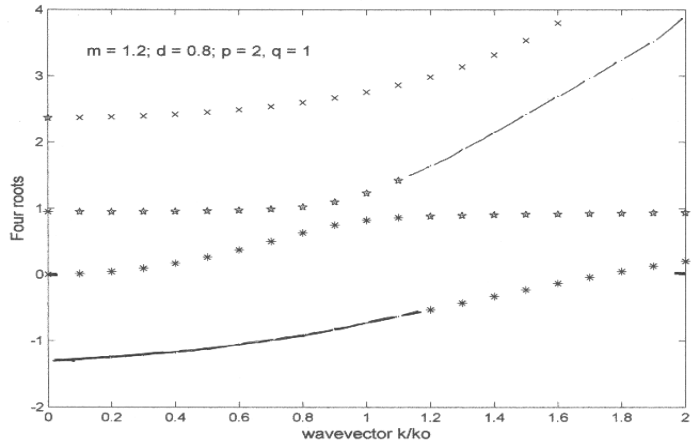

Figure 1:

Shown above are four curves resulting from solution of the quartic of eq.7) with the definitions of eq.8). The four roots are shown as a function of the parameter $k/k_o$, the dimensionless wavevector. There is always a region of spontaneous pattern activation, where one root has a negative region, when $k/k_o$ is below a critical value given by $(k/k_o)_c = q/(md)+p(md)$. The parameter values of the figure are m=1.2, d=0.8, p=2 and q=1

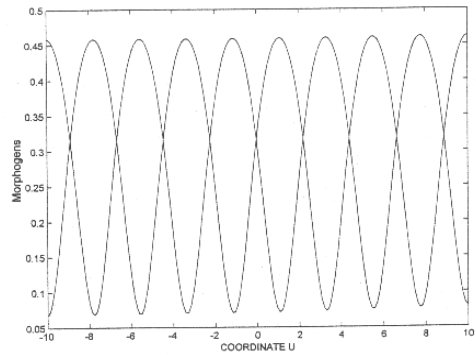

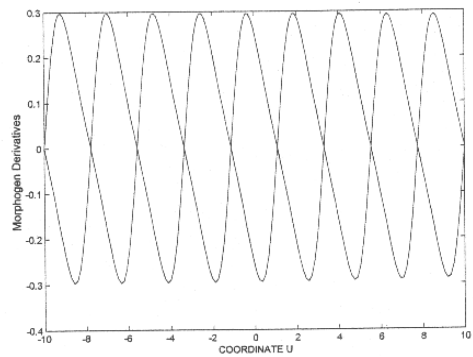

Figure 2:

Shown is a sample numerical solution of the model of eqs.1-5) in the case of constant metric, i.e., on a cylinder in axial symmetry, as a function of the dimensionless parameter 'u'. Both the (dimensionless) morphogens and their derivatives are shown. Solutions shown in all figures are in steady state, where $\partial/\partial t$ has been taken as zero in eqs. 1-5).

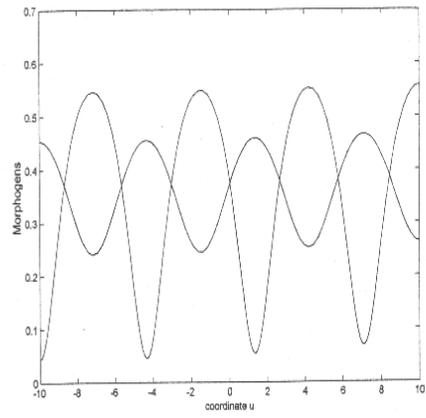

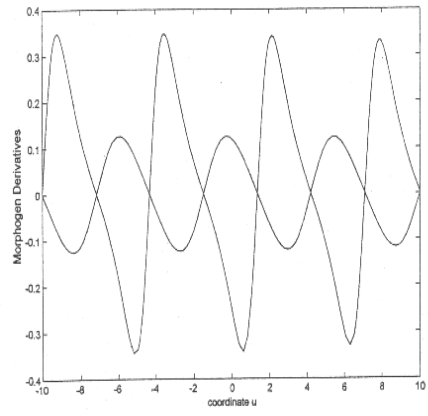

Figure 3:

A sample solution to eqs.1-5) on a sphere is shown, giving both morphogen and derivative. The sphere has a fixed radius and does not respond to the morphogen pattern. The sphere area is here 1.5 times the critical area $A_o$, where $(k_1^2+k_2^2)A_o = 8\pi$. As the radius increases the amplitude of the solutions increase, as the gradients become progressively steeper.

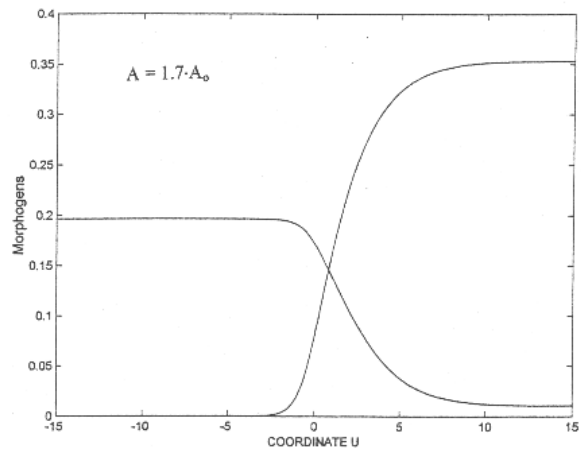

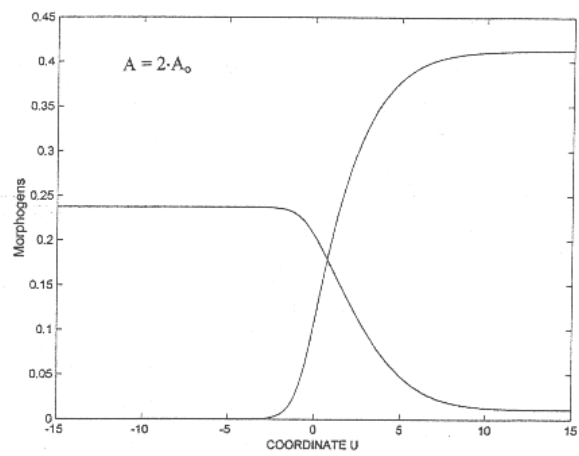

Figure 4:

Two axi-symmetric solutions for each of the two morphogens are shown above as a function of the coordinate 'u'. The coordinate 'u' runs from $-\infty$ to $+\infty$ at the two 'poles', although $-10$ to $+10$ suffices numerically. In contrast to Figure 3, the epithelial shape, via the surface curvatures, is now coupled to the morphogens and vice versa, as discussed in section 4. The two solutions shown correspond to total areas of $A=1.7 \cdot A_o$ and $A=2 \cdot A_o$, where $A_o$ is the sphere radius below which the morphogens are $\equiv 0$. The numerical integration has been carried out for three coupled second order differential equations indicated in the text, and significant surface deformation occurs as the area increases, and growth occurs.

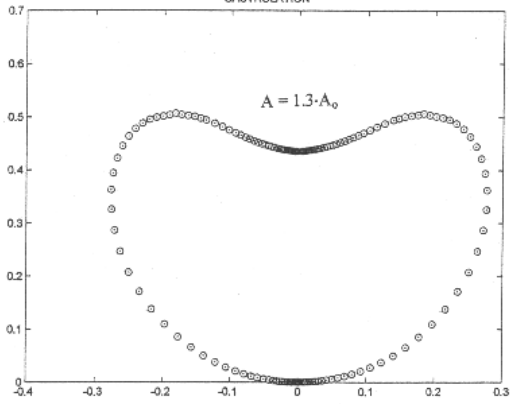
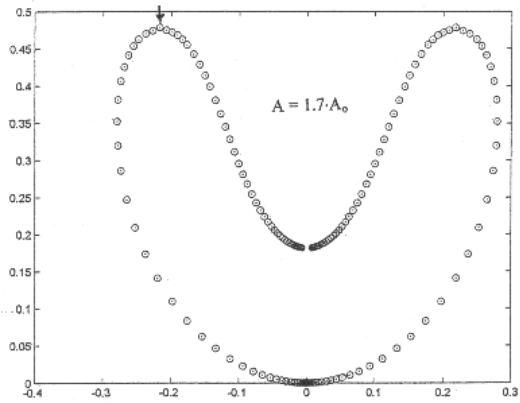
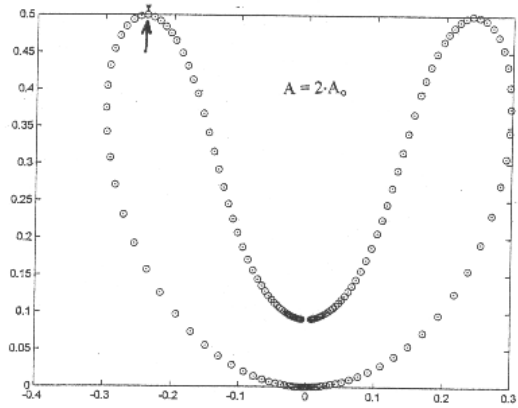

Figure 5:

Shown are the (middle) surfaces in three cases, for total area $A=1.3 \cdot A_o$, $1.7 \cdot A_o$ and $2 \cdot A_o$, the last two areas corresponding to the curves of Figure 4. Growth is simulated by increasing total area, and represents the 'slow' time dependence of the model in contrast to the faster patterning reactions. It is worth noting that at the place denoted by the arrow, the Gauss curvature must become negative. Consideration that any torus (donut) has an integral of K (Gauss curvature) over the donut surface = 0 tells us to expect a Gauss curvature which changes sign at this point. The curvature K crosses zero again to become positive inside the 'gut'. The integral of K over the closed surfaces of the three figures are all given correctly by the Gauss Bonnet theorem as $4 \cdot \pi$. Gastrulation-like movement, resulting in a primitive 'gut', is the most fundamental step in the formation of the first two epithelial types, endoderm and the ectoderm, of any animal. The program integrating the three coupled second order equations is available by email request.